\documentclass[preprint,12pt]{elsarticle}

\usepackage{graphicx}

\usepackage{color}

\usepackage{amssymb}

\journal{Chemical Physics Letter}

\begin{document}

\begin{frontmatter}



\title{Molecular kinetics of solid and liquid CHCl$_3$}


\author[famaf]{Nirvana B. Caballero}
\ead{ncaballe@famaf.unc.edu.ar}

\author[famaf]{Mariano Zuriaga}
\ead{zuriaga@famaf.unc.edu.ar}

\author[qeeri]{Marcelo A. Carignano}
\ead{mcarignano@qf.org.qa}

\author[famaf,qeeri]{Pablo Serra}
\ead{serra@famaf.unc.edu.ar}

\address[famaf]{Facultad de Matem\'atica, Astronom\'{\i}a y F\'{\i}sica, Universidad Nacional de C\'ordoba, C\'ordoba, Argentina and IFEG-CONICET, Ciudad Universitaria, X5016LAE C\'ordoba, Argentina}

\address[qeeri]{Qatar Environment and Energy Research Institute, P.O. Box 5825, Doha, Qatar}

\begin{abstract}
We present a detailed analysis of the molecular kinetics of CHCl$_3$ in a range of temperatures covering the solid and liquid phases. Using nuclear quadrupolar resonance  we determine the relaxation times for the molecular rotations in solid at pre-melting conditions. Molecular dynamics simulations are used to characterize the rotational dynamics in the solid and liquid phases and to study the local structure of the liquid in terms of the molecular relative orientations. We find that in the pre-melting regime the molecules rotate about the C-H bond, but the rotations are isotropic in the liquid, even at supercooled conditions.
\end{abstract}


\end{frontmatter}

\pagebreak


\section*{Introduction}
\label{intro}

Compounds of the form CXCl$_3$  exhibit an interesting phase behavior whose complexity depends on the nature of the X substituent. For example, CBrCl$_3$ and CCl$_4$ have an intermediate plastic phase between the liquid and the pure crystalline phases \cite{Rey:2008ab}. On the other hand, crystalline CHCl$_3$ melts to its liquid form without the intermediate plastic step \cite{FOURME:1966aa}. These compounds, along with other tetrahedral molecules, have received considerable attention in the recent years \cite{CHANG:1995aa,Tironi:1996aa,Llanta:2001aa,Rey:2000aa,Rey:2007aa,Rey:2009aa,Pothoczki:2010aa,Pothoczki:2009aa,Pothoczki:2010ab,Pothoczki:2011aa,Pothoczki:2012aa,Zuriaga:2012aa}. Besides the importance that these molecules have due to their application in several industries and their environmental impact, they represent an interesting academic problem \cite{Rey:2008aa,Romano:2010aa}. In particular, the plastic phases could be regarded as simplified models for liquids that lead to the formation of simplified glass systems. The comparative study of the local order and orientational correlations of these compounds allows us to understand the different microscopic scenarios in the liquid and solid phases  \cite{pardo2007,temleitner2010}.  
In this letter we concentrate our attention on the kinetic and thermodynamic behavior of CHCl$_3$, complementing our previous studies for X=Cl \cite{Zuriaga:2011aa} and X=Br \cite{Caballero:2012aa}. Here we perform a molecular dynamics study (MD) of solid and liquid chloroform, along with nuclear quadrupolar resonance (NQR) measurements of the molecular relaxation times in the solid phase. Our new findings, together with our previous knowledge about the other two systems of the CXCl$_3$ family, allow us to interpret the role of the substituent X in determining the stable phases and their molecular kinetics signature.

\section*{Methods}

NQR experiments were performed on a sample of CHCl$_3$ (Fluka-Sigma-Aldrich \#25670) placed in a glass ampoule and sealed under vacuum.  The $^{35}$Cl NQR transition frequencies and spin lattice relaxation times (T$_1$) were measured between 77 K and melting temperature 210 K by means of a conventional homemade fast Fourier transform (FFT) pulsed spectrometer \cite{suit06}. The relaxation times were measured by using the ($\pi$/2-$\tau$-$\pi$/2) sequence. The temperature was controlled within 0.1 K using a homemade cryostat.

We performed MD simulations of {CHCl$_3$} in solid and liquid conditions in the NPT ensemble using the Gromacs v.4.5.5 simulation package \cite{Hess:2008aa}. The dynamics equations were integrated using the leap-frog algorithm with a time step of 0.001 ps. The simulated system consisted of 3920 molecules (19600 atoms), periodic boundary conditions were applied in all three Cartesian directions and a spherical cutoff was imposed at 2 nm for all interactions. The temperature was controlled with the velocity rescale algorithm \cite{Bussi:2007aa} with a time constant of 0.2 ps, and the pressure maintained at 1 atm with a Berendsen anisotropic barostat with a time constant of 0.5 ps and compressibility of 7 $\times 10^{-6}$ bar$^{-1}$. 

The MD simulations of the solid phase of CHCl$_3$ were performed using as initial configuration the structure determined by X-ray diffraction\cite{FOURME:1966aa}. The crystalline structure of chloroform is orthorhombic and corresponds to the space group Pnma 62, with four molecules per unit cell. The melting temperature at ambient pressure is 210 K. The lattice parameters, which were determined at 185 K, are $a=7.485$ \AA, $b=9.497$ \AA\ and $c=5.841$ \AA\ \cite{FOURME:1966aa}. The initial configurations for the simulations in the solid phase consist of perfect crystalline systems of $10 \times 7 \times 14 \,=\,980$  unit cells, which include $980 \times 4=3920$ molecules. These configurations were relaxed at each temperature with short NVT runs of 100 ps.

Several models for CHCl$_3$ consisting of different sets of Lennard-Jones parameters, charges and slightly different interatomic distances can be found in the literature \cite{DIETZ:1984aa,JORGENSEN:1990aa,KOVACS:1990aa,Hubel:2006aa}, and a comparison between  them and several parameter variations can be found in Ref. \cite{Martin:2006aa} for liquid chloroform at ambient conditions. Here, after tests in the liquid and solid phases, we adopted one variation of a model originated by Dietz and Heinzinger \cite{DIETZ:1984aa} with atomic charges resulting from an optimized HF/aug-cc-pVDZ geometry \cite{Martin:2006aa}. This five sites CHCl$_3$ model is based on a rigid, distorted tetrahedron with the C atom at the center, three Cl atoms on three vertices, and a H atom located in the remaining vertex. The intermolecular interactions were described by a Lennard-Jones (L-J) potential plus a Coulombic term. The L-J parameters ($\sigma$,$\epsilon$) for interactions between different atoms were determined using the combination rules $\sigma_{ij}=(\sigma_i + \sigma_j)/2$ and $\epsilon_{ij}=\sqrt{\epsilon_i\epsilon_j} $, where $i$, $j$ represent the three atomic species  C, Cl and H. The details of the parameters and model architecture are summarized on Table \ref{tabla1}.

\begin{table}[!ht]
\caption{\label{tabla1}Model parameters for the {CHCl$_3$} molecule.}   
\centering                                 
\begin{tabular}{l | c c c}
\hline\hline                     
Atom & $\sigma$ [nm] & $\epsilon$ [kJ/mol] & $q$ [e]\\                      
\hline 
C    & 0.340 & 0.427  &  0.569 \\
Cl   & 0.344 & 1.255  & -0.177 \\H    & 0.280 & 0.084  & -0.038 \\
\hline \hline 
\multicolumn{4}{c} {$d$(C-Cl)=0.1769 nm} \\
\multicolumn{4}{c} {$d$(C-H)=0.1078 nm} \\
\multicolumn{4}{c} {$d$(Cl-H)=0.2337 nm}\\           
\hline \hline 
\end{tabular}
\end{table}

\section*{Results}

\begin{figure}[!t]
\centering
{\includegraphics[width=0.7\textwidth]{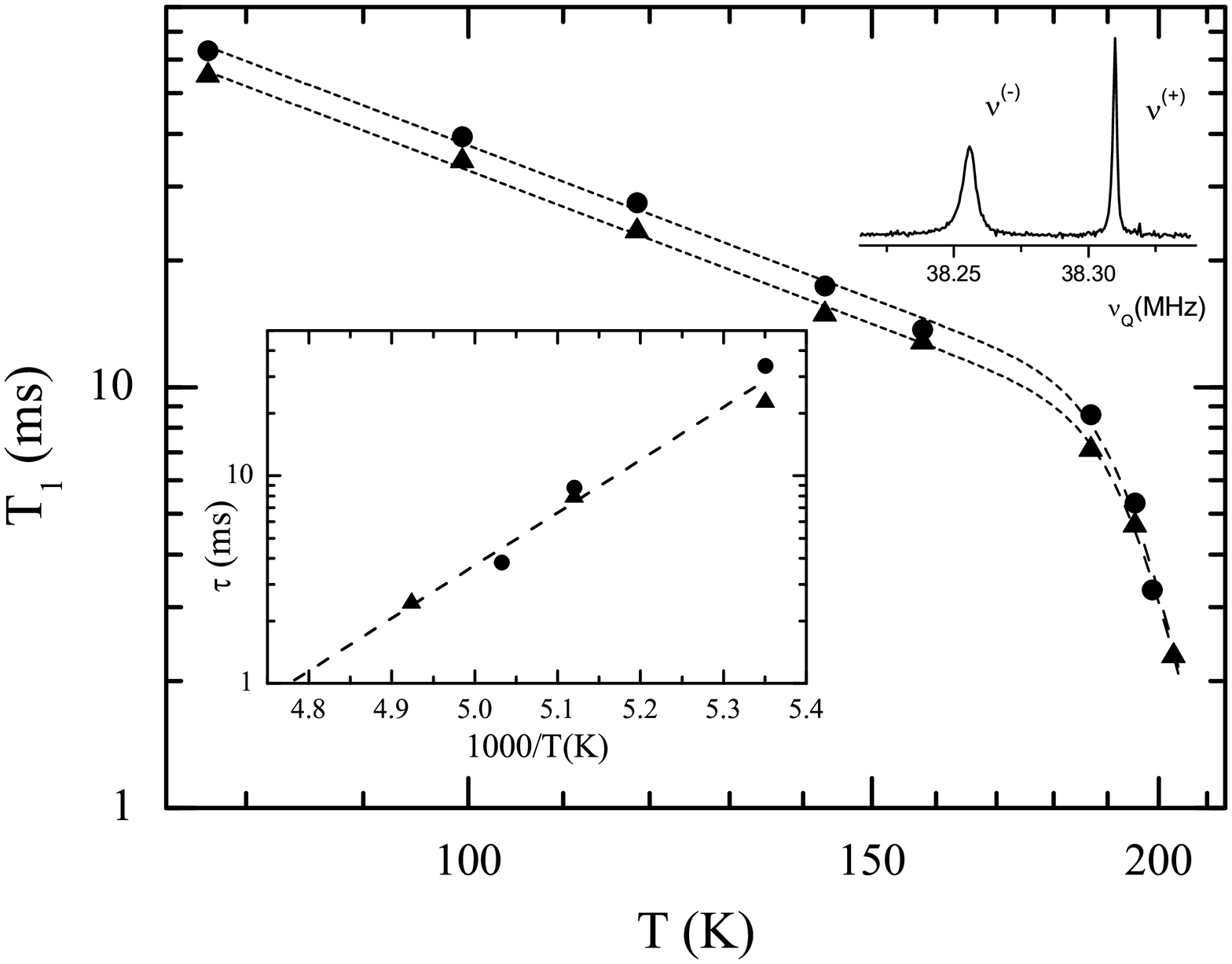}}
\caption{\label{nqr}  $^{35}$Cl Spin-lattice relaxation time of $\nu^{(+)}$ (circles) and $\nu^{(-)}$ (triangles). Upper inset shows the NQR spectra at 77 K. Lower shows inset reorientational correlation time $\tau$ extracted from both $T_1$ data.}
\end{figure}

In order to determine the correlation times corresponding to the rotational jumps of CHCl$_3$ we measured the NQR spin lattice relaxation times, $T_1$, between 77 K and the melting point at 210 K. In the solid phase, the CHCl$_3$ spectra displays only two NQR lines with an intensity ratio of 2:1, see top-right inset of Figure \ref{nqr}. These two lines correspond to the the two nonequivalent Cl atoms in the asymmetric unit as revealed by the crystallographic studies of Fourme and Renaud \cite{FOURME:1966aa}. The NQR relaxation times of the two lines are shown in Figure \ref{nqr}. 

Below 180 K the $T_1$ data shows a weak temperature dependence ($T^{-2}$) while above that temperature $T_1$ has the behavior characteristic of thermally activated process. In molecular crystals and for rigid molecules, the quadrupolar relaxation is usually due to two types of mechanisms: a phonon or molecular libration contribution \cite{WOESSNER:1963aa} and a rotational contribution resulting from the slow reorientational jumps of the molecules between equivalent positions \cite{Alexander:1965aa}. Then, the relaxation rates for each nucleus that participates in the reorientation can be described by the sum of these two independent processes, and for a three-fold symmetric molecule \cite{Zuriaga:2012aa} $T_1^{-1}=aT^2+3/4 \tau^{-1}(T)$. Therefore, subtracting the phonon contribution, the reorientational contribution to the relaxation process, $\tau^{-1}(T)$, can be extracted from the data and the resulting  reorientational correlation time $\tau$ is shown in the bottom-left inset of Figure \ref{nqr}.

We began our MD study by estimating the melting temperature of the model. 
In order to have an estimation of the model's melting temperature we performed several tests, starting from checking the stability of the crystalline phase as a function of the temperature. It was found that at 240 K the system remains as a solid for up to 50 ns of simulation. At 250 K the system melts in less than 1.5 ns. Since a small crystalline sample such as the simulated one is expected to persist as a solid under superheated conditions, 250 K represents an upper limit for the melting temperature of the model CHCl$_3$. Next we study the stability of a system consisting of a solid slab in contact with a liquid slab, and it was found that the system melts at 180 K but remains stable for up to 50 ns at 160 K and lower temperatures. In this case, the later temperature represents a lower bound for the melting temperature since the fluctuations, which are relatively large due to the intrinsic small size of the simulated system, tend to favor the phase with larger entropy. Therefore, the model's melting temperature is between these two limits, and although this is a large interval, it suffices for the purpose of the analysis that we present below.

\begin{table}[!t]
\centering                                 
\begin{tabular}{l | c c c c}
\hline\hline                     
T [K] & $k_R$ & $P(k_R)$ & $\tau_{Cl}$ [ps] \\
\hline 
190    & 0      & 0.822        & -- \\
200    & 1      & 0.161        & -- \\
205    & 2      & 0.016        & -- \\
210    & 10     & $10^{-14}$   & $1.5 \times 10^7$ & \\
215    & 33     & $10^{-61}$   & $4.7 \times 10^6$ & \\
220    & 72     & $10^{-156}$  & $3.0 \times 10^6$ & \\
230    & $>$100 & $<10^{-230}$ & $5.1 \times 10^5$ & \\
240    & $>$100 & $<10^{-230}$ & $8.9 \times 10^4$ & \\
\hline \hline 
\end{tabular}
\caption{\label{tabla}Molecular rotations in solid {CHCl$_3$}. $k_R$ is the observed number of molecules that rotated at least once in a simulation of $t_0=50$ ns, $P(k_R)$ is the corresponding probability according with the experimental relaxation time (see text) and $\tau_{Cl}$ is the relaxation time of the C-Cl molecular bonds calculated from the MD simulations.}
\label{rots}
\end{table}

\begin{figure}[!t]
\centering
{\includegraphics[width=0.5\textwidth]{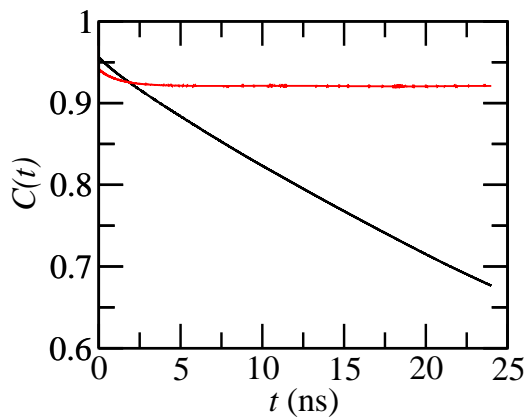}}
\caption{\label{acfs} Rotational auto correlation function of the C-Cl (black) and C-H (red) bond vectors for an {\em overheated} solid of CHCl$_3$ at 240 K.}
\end{figure}

The NQR measurements show that the orientational relaxation time is $\tau^\ast=1$ ms at 205 K. This time is essentially infinity for an atomistic simulation so we cannot perform a direct comparison between experiments and simulations. However, in simulations of 50 ns we are able to see some molecular reorientations that become more frequent as the temperature is increased. These molecular motions are in all cases sudden jumps between equivalent positions achieved by rotations of 120$^\circ$ about the C-H bond. As a result, the auto correlation function of the C-H bonds does not show any decay with time besides a sharp small drop at very short times due to the molecular librations. The sudden 120$^\circ$ rotations are reflected by a slow decay of the auto correlation function of the C-Cl bonds, see Figure \ref{acfs}. Then, we analyze the rotational jump dynamics in a total time of $t_0=50$ ns. This is done assuming that the rotations of a single molecule follow a Poisson distribution characterized by the experimental rate of rotations, {\em i.e.} $\lambda=5 \times 10^{-5} (= t_0/\tau^\ast)$. The probability that a molecule jumps at least once, $p=1-e^{-\lambda}=0.00005$, is taken as the probability of success in a binomial distribution considering $N=3920$ independent trials. Next we calculate the probability of having $k_R$ molecules jumping at least once during the simulated time $t_0$, which is shown in Table \ref{rots}. The table also includes the relaxation times calculated from the rotational autocorrelation functions of the C-Cl bond vector, denoted by $\tau_{Cl}$. These correlation times, much longer than the total simulation time, were estimated from the initial slope of autocorrelation functions assuming an exponential decay. The expectation value for the number of molecules having at least one rotation is $Np=0.196$, which approximately coincides with the number of molecules that jump at least once observed in the simulation at 200 K. Moreover, as the temperature increases above 210 K the number of molecular jumps dramatically increases and therefore cannot be explained in terms of the experimental relaxation time at pre-melting conditions. This collection of results, the lower and upper bounds described above together with the probability analysis, point to a melting temperature of the simulated model very close to the experimental melting temperature of 210 K.

\begin{figure}[!h]
\centering
{\includegraphics[width=0.5\textwidth]{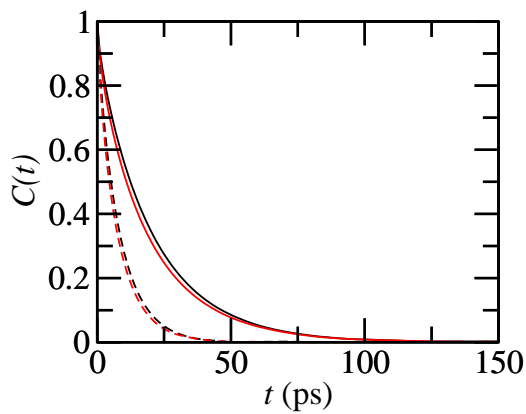}}
\caption{\label{acfl} Rotational auto correlation function of the C-Cl (black) and C-H (red) bond vectors for liquid CHCl$_3$ at 190 K (solid lines) and 240 K (dashed lines).}
\end{figure}

The molecular rotations in the solid phase are essentially about the C-H bond vector. This is clear from the auto-correlation functions of the C-Cl and C-H bond vectors displayed in Figure \ref{acfs} for the {\em overheated} solid at 240 K. Both correlations show a fast initial decay from 1 to $\sim$0.95 due to the molecular librations, but the C-H bond remains in the same direction while the C-Cl bonds relax due to the sporadic molecular jumps. The question that arises is whether this anisotropy in the rotations is preserved upon melting or not. In Figure \ref{acfl} we show the corresponding curves for a liquid system at a stable 240 K, and at {\em supercooled} conditions at 190 K. The two systems show a similar rate for the rotations about the two distinct bond vectors. The disappearance of the rotational anisotropy is possible due to a decrease in the density of the system of approximately 10 \% as it melts. 

\begin{figure}[!t]
\centering
{\includegraphics[width=0.5\textwidth]{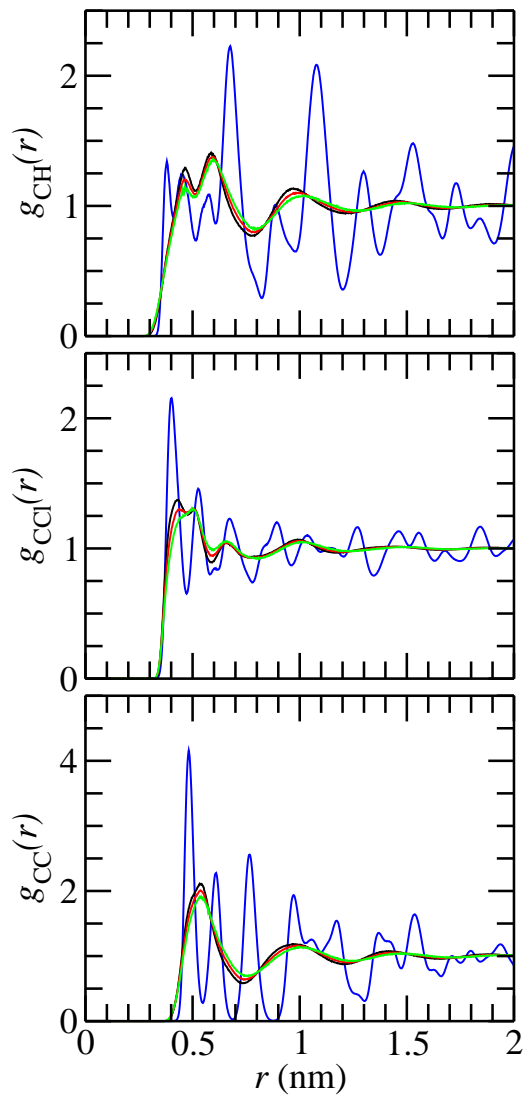}}
\caption{\label{gofr} Pair distribution functions for the CH (top panel), CCl (center panel) and CC (bottom panel) pairs. The black, red and green lines correspond to the liquid system at 190 K, 240 K and 300 K, respectively. The blue line corresponds to the solid phase at 190 K.}
\end{figure}

In order to understand what are the molecular rearrangements upon melting that lead to the decrease in density we analyze the pair distribution functions, $g(r)$, for the different atom pairs. In particular, we plot on Figure \ref{gofr} the $g(r)$ for C-H and C-Cl pairs. The curves corresponding to the solid phase at 190 K display the signature multi-peak pattern of the lattice structure. The pair correlation functions for the liquid systems reveal the local structure of the fluid that dies off for $r \geq 1.5$ nm. The picture that emerges is that, as the system melts, the strong C$\cdots$H association in the solid phase is lost resulting in an overall expansion of the volume. This is clear from $g_{CH}(r)$ that has a distinctive peak at $r=0.38$ nm in the solid structure that completely disappears in the liquid, moving the peak to $r=0.46$ nm. The CC pair correlation functions also reflect the overall expansion of the volume, which in turn, provides sufficient space for the molecules to rotate with no preferential axis. 
The strong C$\cdots$H association is the result of the electrostatic interactions at short distances, which has also been considered to be the microscopic origin of polar phases in chloroform and bromoform at higher pressure \cite{dziubeck2008}.

\begin{figure}[!t]
\centering
{\includegraphics[width=0.5\textwidth]{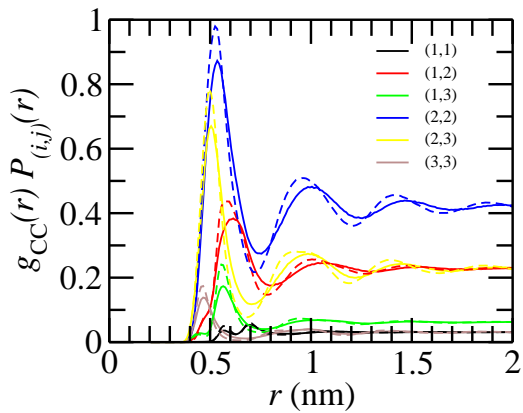}}
\caption{\label{rey} Probability of finding different relative orientations between neighboring molecules at 190 K (dashed line) and 300 K (solid line).}
\end{figure}

A more detailed analysis of the local structure of the liquid, based on a geometrical criteria, has been applied recently to several tetrahedral molecular systems of the form ZY$_{i,4}$, where the four Y atoms may be of different type \cite{Rey:2008aa}. The idea is based on considering the number of Y atoms between the two parallel planes containing the Z atoms of neighboring molecules. Depending on this number and to which molecules these Y atoms belong, six different classes of molecular arrangements can be defined: $(1,1)$, $(1,2)$, $(1,3)$, $(2,2)$, $(2,3)$ and $(3,3)$. The numbers refer to the number of atoms from each molecule located between the two parallel planes. The probabilities of these different classes, given that the distance between the two Z atoms is $r$, is denoted by $P_{(i,j)}(r)$. In a recent study of CBrCl$_3$ \cite{Caballero:2012aa} we claimed that a more meaningful quantity is the product $g_{CC}(r) \times P_{(i,j)}(r)$, which produces partial radial distribution functions, since it directly removes those configurations that have negligible contribution to the overall structure of the system. In Figure \ref{rey} we show $g_{CC}(r) \times P_{(i,j)}(r)$ for the liquid system at 190 K and 300 K. There are no major differences for the two cases, only the expected softening of the liquid structure as the temperature is increased. At short distances, the {\em edge-to-edge} and {\em edge-to-face} configurations are the most probable. {\em face to-face} arrangements are meaningful at very close distances, and {\em corner-to-face} and {\em corner-to-edge} have their maximum contribution at slightly larger separation between molecules. This general pattern and relative order of the different molecular orientations is remarkably similar to one observed for the case CCl$_3$Br, that is, with a large substituent atom.

\begin{figure}[!t]
\centering
{\includegraphics[width=0.5\textwidth]{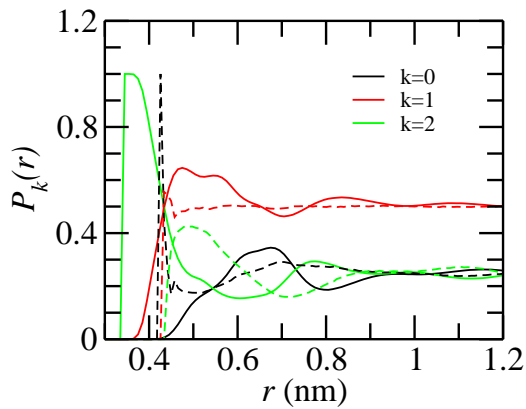}}
\caption{\label{reyka} Probability of finding different relative orientations between neighboring molecules for CHCl$_3$ (solid line) and CBrCl$_3$ (dashed line) in the liquid phase at 300 K.}
\end{figure}

\begin{figure}[!t]
\centering
{\includegraphics*[width=0.5\textwidth]{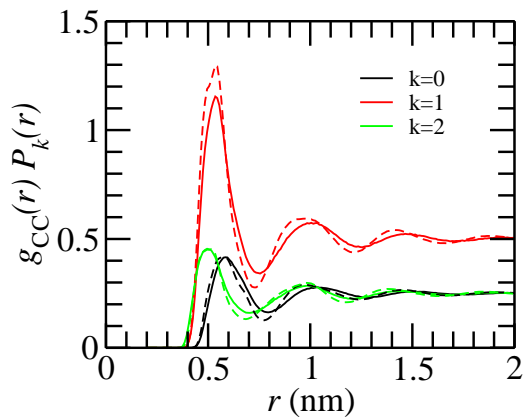}}
\caption{\label{reyk} Product of the conditional probabilities $P_k$ times the C-C radial distribution function for CHCl$_3$ at 190 K (dashed line) and 300 K (solid line).}
\end{figure}

Further insight on the local molecular ordering can be gained by looking at the probability $P_k$ of finding $k$ substituent atoms ($k=0$, $1$ or $2$) between the two parallel planes defined by the C atoms of neighboring molecules. Moreover, the size of the substituent atom may play a role in this probability and therefore we compare our new results for CHCl$_3$ with our previous calculations for CBrCl$_3$. In Figure \ref{reyka} we show $P_k$ for the two systems at 300 K. While for a H substituent at closest approach $P_2$ is maximum, for the Br case is $P_0$ the most probable configuration. This can be understood from pure geometrical considerations. Since the H atom is much smaller than the Cl atoms, the closest approach between two molecules happens with a {\em corner-to-corner} configuration directly placing the H atoms facing each other. Since the Br is considerably larger than the Cl atom the closest approach results from a {\em face-to-face} arrangement with the Br atom on the opposite sides of the facing planes. Interestingly, this difference is washed away when we introduce the $g_{CC}(r)$ as a weighting factor, as shown in Figure \ref{reyk} for the liquid system at 190 K and 300 K, which shows remarkable similarity with the corresponding curves for CBrCl$_3$. For large separations, the probabilities go to 0.5 for $k=1$ and 0.25 for $k=0$ and $k=2$, which are the expected values from the relative number of the different atomic species in the system. At short range, the most probable configurations are those with $k=1$. 

A comparative analysis of the findings of this work with those corresponding to CCl$_4$ and CBrCl$_3$ allow us to draw some general conclusions. The essential difference among these three molecules is the size of the substituent atom. The perfect tetrahedral structure has an intermediate fcc plastic phase between the solid and the liquid. Substituing a Cl atom by a larger Br atom preserve the intermediate plastic phase. However, the substitution with a much smaller H atom suppresses completely the plastic phase. The dynamics of the CHCl$_3$ solid phase is {\em qualitatively} different than that of the other two compounds: while rotations about all the bonds are observed in the CBrCl$_3$ and CCl$_4$, only rotations about the C-H bond are possible in the present case. 

\section*{Acknowledgement}
NB, MZ and PS and would like to acknowledge SECYT-UNC, CONICET, and MINCyT C\'ordoba for partial financial support of this project.


\pagebreak

\end{document}